\documentclass{revtex4-1}

\newcommand{\xx}[1]{#1}

\usepackage{amsmath}    
\usepackage{graphicx}   
\usepackage{verbatim}   
\usepackage{color}      
\usepackage{subfigure}  
\usepackage[plainpages=false]{hyperref}   
\usepackage{cleveref}

\begin{document}



\title{Linear rheology of reversibly cross-linked biopolymer networks}

\author{Henry Amuasi, Andreas Fischer, Annette Zippelius and Claus Heussinger}
\affiliation{Institute of Theoretical Physics, Georg-August University of G\"ottingen, 37073 G\"ottingen, Germany}


\date{\today}

\begin{abstract}
  We suggest a simple model for reversible cross-links, binding and
  unbinding to/from a network of semiflexible polymers. 
  The resulting frequency dependent response of the network to an
  applied shear is calculated via Brownian dynamics simulations. It is
  shown to be rather complex with the timescale of the linkers
  competing with the excitations of the network. If the lifetime of
  the linkers is the longest timescale, as is indeed the case in most
  biological networks, then a distinct low frequency peak of the loss
  modulus develops. The storage modulus shows a corresponding decay
  from its plateau value, which for irreversible cross-linkers extends
  all the way to the static limit. This additional relaxation
  mechanism can be controlled by the relative weight of reversible and
  irreversible linkers.
 \end{abstract}

\pacs{}

\maketitle


\section{Introduction}

The simulation of filamentous polymer networks (e.g. cytoskeletal
networks) presents a substantial challenge as to the broad spectrum of
length-scales and relaxation times~\cite{Unterberger2014}. At high
frequencies the response is governed by small-wavelength bending
fluctuations of individual filaments. The higher the driving frequency
the smaller the dominant wavelength, which leads to the well-known
frequency-dependence for the modulus $g\sim \omega^{3/4}$
\cite{koenderink2006,gittes1998}. At intermediate frequencies
collective network modes come into play and the response is mainly
elastic~\cite{lieleg2007b}. Different theories have been devised to
understand this
regime~\cite{heussinger2007,mackintosh1995,broedersz2014}.  The low
frequency regime of the modulus is sensitive to the dynamics of the
cross-linking proteins. If these provide permanent connections and are
sufficiently numerous for the network to percolate, then the low
frequengy limit is characterized by a finite elastic modulus 
~\cite{gardel04:_elast_behav_cross_linked_bundl_actin_networ}. However,
crosslinks in biological networks usually have a lifetime
$\tau_{\rm cl}$ of only several seconds
\cite{lieleg2009,Ehrlicher6619}, so that the network can undergo a
terminal relaxation at frequencies $\omega < 1/\tau_{\rm cl}$. In this
low-frequency regime the network flows like a liquid, e.g. governed by
filament reptation and constraint release~\cite{lang18:_disen} in
entangled networks, or repeated crosslink un/rebinding processes in
crosslinked networks
\cite{ward08:_dynam_viscoel_actin_cross_linked,broedersz2010,lieleg2008}.

While there are ample simulations dealing with intermediate and high
frequencies
\cite{huisman2008,huisman2010,amuasi2015,head2003,PhysRevLett.95.178102,kim09:_comput},
efforts to combine the entire frequency range in one simulation are
scarce \cite{cyron13:_microm,PhysRevLett.112.238102}. Here we present
an alternative approach. With the high-frequency branch being well
understood, we sacrifice high-frequency bending fluctuations of
individual filaments, treating the filaments with the help of an
effective potential of mean force. A similar approach was first
described in two-dimensional Mikado networks in
Ref.~\cite{heussinger2007_2}. Our three-dimensional simulations are
built on the method of Huisman {it et al.} \cite{huisman2008} and
Amuasi {\it et al.} \cite{amuasi2015}.
Within this approach we study in detail the process of
reversible crosslinking and its role for the rheological response of
the filament network to small angle oscillatory shear.

\section{Model}

We study the frequency dependent visco-elastic response of a randomly
cross-linked biopolymer network. It is our aim to work out and
understand the differences between reversibly and irreversibly-bound
cross-links. To this end we simulate the Brownian dynamics of randomly
cross-linked filament networks, treating the positions of the
cross-links as dynamical variables.

The polymer segments represent connections between neighboring
crosslinks and thus mediate interactions between them. We ignore the
precise configurations of the polymer segments and instead work with
effective, spring like interactions between the cross-links. This
approximation is well justified for low frequencies, where the short
wavelength modes of the polymer segments are relaxed.

To account for reversible crosslink binding we also allow the polymer
segment length in between cross-links to vary. In this model a
crosslink unbinding/rebinding event is treated as diffusion of the
crosslink along the filament. Polymer-mediated forces acting on the
crosslink then act as bias to this diffusion process. \xx{This type of
  description is useful to understand the limiting case of fast un-
  and rebinding. The opposite limit of slow binding has to be dealt
  with by stochastic transitions modeled with the help of appropriate
  Metropolis Monte Carlo steps.}

We model the effective polymer-mediated interactions between the
cross-links at positions $\{{\bf r}_i\}_{i=1}^N$ by
\begin{equation}\label{eq:H}
H=\sum_{ij}k_2^{(ij)}(|{\bf r}_{ij}|-l_{(ij)})^2+\sum_{ijk}k_3^{(ijk)}\theta_{ijk}^2
\end{equation}
The first term represents the polymer stretching energy with stiffness
$k_2^{(ij)} = \alpha/l_{ij}$. Here $l_{ij}$ is the contour length
between cross-link $i$ and $j$.  The parameter $\alpha$ is a constant
that, in the context of athermal beam stretching, takes the meaning of
Youngs modulus multiplied by cross-sectional area. The second term in
Eq.~(\ref{eq:H}) represents the polymer bending energy, restraining
the angle, $\cos\theta_{ijk}=\hat{{\bf r}}_{ij}\cdot \hat{{\bf r}}_{jk}$, between
any three consecutive crosslinks $ijk$ along the same polymer. The
bending stiffness is taken to be
$k_3^{(ijk)}=\frac{3}{2}k_B Tl_p/(l_{ij}+l_{jk})$. The scale for the
bending stiffness is set by temperature $k_BT$ and the persistence
length of the polymer $l_p$. For more information on the model
Hamiltonian, see reference \cite{amuasi2015}.

We assume the network to be embedded in a viscous fluid of viscosity
$\eta$, giving rise to viscous drag and thermal noise. In the
overdamped limit the cross-links perform Brownian motion, described by
a Langevin equation:
\begin{equation}
\eta\dot{{\bf r}}_i={\bf F}_i+{\pmb \xi}_i.
\label{eq_cLPosition_Final}
\end{equation}
The systematic force is given by
${\bf F}_i=\partial H /\partial{\bf r}_i $ and the noise is chosen
according to the FDT with zero mean and variance
$\langle{\pmb \xi}_i(t)\cdot {\pmb \xi}_j(t^\prime)\rangle=2k_B
T\eta\;\delta_{ij}\delta(t-t^\prime)$.

Physiological crosslinking proteins can unbind thermally or under the
application of mechanical forces, since the bonds they form are rather
weak. Here we present a model for the crosslink unbinding process that
incorporates both thermal and forced unbinding. \xx{In both cases, we
  assume that --after unbinding-- the crosslink immediately rebinds to
  the filament, possibly at another nearby location. In this
  ``fast-rebinding'' limit we arrive at a description of crosslink
  binding in terms of a biased diffusion process.}

\begin{figure}[h!tbp]\centering
\includegraphics[width=0.9\columnwidth]{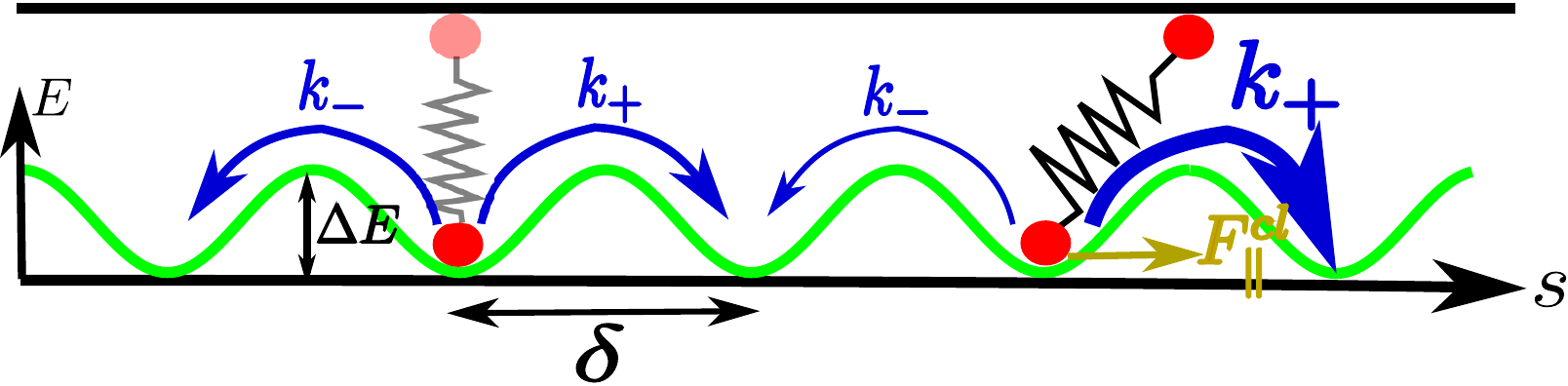}
\caption{Scheme of the unbinding process: For a crosslink (depicted as
  a spring with two heads), a filament corresponds to a periodic
  energy landscape with periodicity $\delta$ (green) along the
  filament's arclength $s$. Energy minima represent crosslink binding
  sites. Unbinding corresponds to a jump over the energy barrier
  $\Delta E$ to the next minimum. Left: symmetric jump rates $k_+$,
  $k_-$ (force-free), right: asymmetric rates (due to the force
  $F_{\|}^{cl}$ from the spring)}
\label{fig:crosslink1}
\end{figure}

A crosslink may be regarded as a spring with two heads, each of which is
binding to a different filament (see Fig.~\ref{fig:crosslink1}). With
respect to crosslink binding, a filament can be considered as a
periodic energy landscape along its arc length $s$, the minima of
which represent the binding sites. The double-helical shape of F-actin
suggests a periodicity of $\delta \approx 50\,$nm. We model a thermal
unbinding event as an activation process with rate
$k=\tilde{k}_0 e^{-\beta\Delta E} $.

Remembering that the heads of the crosslink are coupled via a spring,
one has to account for the additional force ${\bf F}^{cl}$ on a head,
arising when the spring is stretched or compressed.  We assume that
the unbinding kinetics is determined by the component $F_{\|}^{cl}$
parallel to the local tangent of the polymer: the jump rate in the
direction of $F_{\|}^{cl}$ is enhanced and decreased in the opposite
direction (see Fig.~\ref{fig:crosslink1}), which breaks the symmetry
of the force-free case. Assuming $F_{\|}^{cl}$ is pointing to the
right, this changes the jump rates to
\begin{align}\label{jumprates}
k_+=\tilde{k}_0 e^{-\beta(\Delta E - F_{\|}^{cl} \delta)} , \,\,\,\, k_-=\tilde{k}_0 e^{-\beta(\Delta E +F_{\|}^{cl} \delta)}
\end{align}
The rate-asymmetry leads to an effective movement of the
crosslink-head along the polymer, which can be described by 
a Master equation for the probability $P_s(t)$ to find the crosslink at binding site $s$ at time $t$:
\begin{equation}
\partial_tP_s(t)=k_+P_{s-1}(t)+k_-P_{s+1}(t)-(k_++k_-)P_s(t)
\end{equation}
Multiplying the above equation by $s$ and summing over all $s$, we find an equation for the average velocity of the crosslink
\begin{equation}
v_s:=\partial_t\sum_ss P_s(t)\delta=2k_0\sinh(\beta\delta F_{\|}^{cl} )
\end{equation}
wiht $k_0=\tilde{k}_0 e^{-\beta(\Delta E)}$.
Assuming  a small pulling force, so that linear response
applies, the equation of motion reads explicitly
\begin{align}
\zeta v_s = F_{\|}^{cl} + \xi 
\label{equ:clMov_equ}
\end{align}
Here $\zeta=k_B T/(2k_0\delta^2)$ denotes the mobility, related to the
diffusion constant $D_{cl}=2k_0\delta^2$ in the usual way, and we have
added a noise term with $\left<\xi(t)\right>=0$ and
\begin{equation}
\left<\xi(t) \xi(t_0)\right>=2 k_B T\zeta\,\delta(t-t_0).
\end{equation}

\begin{figure}[h!tbp]\centering
  \includegraphics[width=0.5\columnwidth]{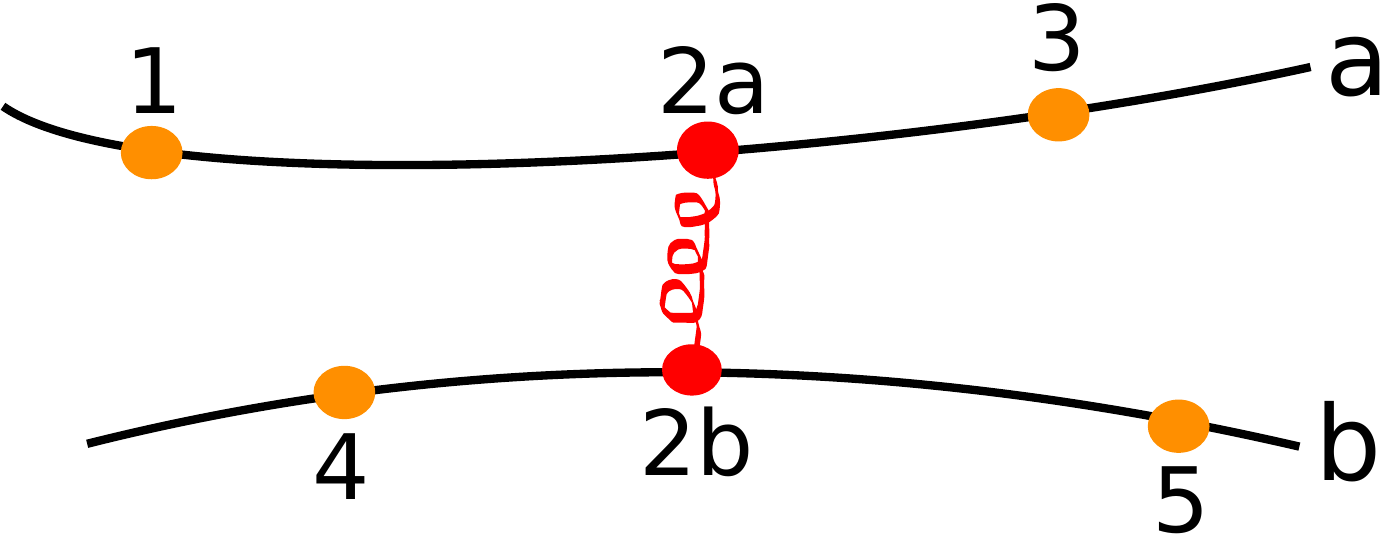}
  \caption{Crosslink modelled as a spring embedded in a network with
    adjacent crosslinks (pointlike), labeled 1,3,4,5.}
  \label{fig:crosslink_inside_network}
\end{figure}

The spring is stretched or compressed due other cross-links connecting
to the two filaments under consideration (see
Fig.\ref{fig:crosslink_inside_network}). The forces acting on the 2
heads of the motor at positions ${\bf r}_{2a}$ and ${\bf r}_{2b}$
\begin{align}\label{eq:two_heads}
\eta\dot{{\bf r}}_{2a}&={\bf F}_{2a}+{\bf F}^{cl}\nonumber\\
\eta\dot{{\bf r}}_{2b}&={\bf F}_{2b}-{\bf F}^{cl}
\end{align}
are decomposed into the forces due to the other cross-links,
${\bf F}_{2a}$ and ${\bf F}_{2b}$, and the force, ${\bf F}^{cl}$, due to the
springlike cross-link, considered explicitly. 

For the simulations we consider the crosslink as point-like, thus the
two heads of the crosslink are at identical positions, which can be
achieved approximately by a high spring constant for the crosslink. In
this limit the two heads move with the same velocity, so that
Eqs.~(\ref{eq:two_heads}) can be solved for
${\bf F}^{cl}=({\bf F}_{2b}-{\bf F}_{2a})/2$. The equations of motion
for the two heads of the crosslink connecting filaments $a$ and $b$
then read:
\begin{align}
\label{arclength_dynamics}
\zeta v_{2a}&=F_{\|}^{a}+\xi_a\\
\zeta v_{2b}&=F_{\|}^{b}+\xi_b
\end{align}
giving rise to a change in arclength
$\dot{l}_{12}=-\dot{l}_{23}= v_{2a}$ and
$\dot{l}_{42}=-\dot{l}_{25}= v_{2b}$. Since we do not keep track of
the contour of the polymer between cross-links, we have approximated
the local tangent as
$F_{\|}^{a}={\bf F}^{cl}\cdot ({\bf r}_3-{\bf r}_1)/|{\bf r}_3-{\bf
  r}_1|$ and correspondingly for $F_{\|}^{b}$.

In turn, a changing arclength $l_{(ij)}$ modifies the spring constants
in Eq.~(\ref{eq:H}) and thus the forces acting on the crosslinks. This
mechanism provides the coupling between the spatial (lab-frame)
degrees of freedom ${\bf r}_i$ of the crosslinks, and the internal
degrees of freedom $s_a$, measured by the position along the filament.

\subsection{Technical details}

As units we choose $l_p$, $k_BT$ and $\eta$. In these units, the
parameter $\alpha$ is taken as $\alpha=349$, which is motivated by the
comparison of a spring constant of a beam with that of a wormlike
chain.

We simulate 1000 crosslinks on 300 filaments each of length $1.28$ in
a simulation box, which is taken to be of length $1$. Periodic
boundary conditions are assumed. To incorporate shear deformations we
use the method of Lee and Edwards.
To measure the frequency-dependent modulus we apply a shear strain
$\gamma =\gamma_0\sin(\omega t)$ frequency $\omega$ and with
$\gamma_0 = 0.008\ldots 0.16$ depending on frequency. The resulting
stress is fitted to the form
$\sigma = \gamma_0(G'\sin(\omega t)+G''\cos(\omega t))$, which defines
real and imaginary part of the complex modulus
$G(\omega)=G'(\omega)+iG''(\omega)$. In order to prevent two
crosslinks to collapse into a single point we implemented a minimum
distance $l_{\rm min}=0.1l_p$ between two neighboring crosslinks on a
filament via the repulsive part of a Lennard-Jones potential.

\section{Results: Irreversible cross-links}

\begin{figure}[tbp]\centering
  \includegraphics[scale=0.24]{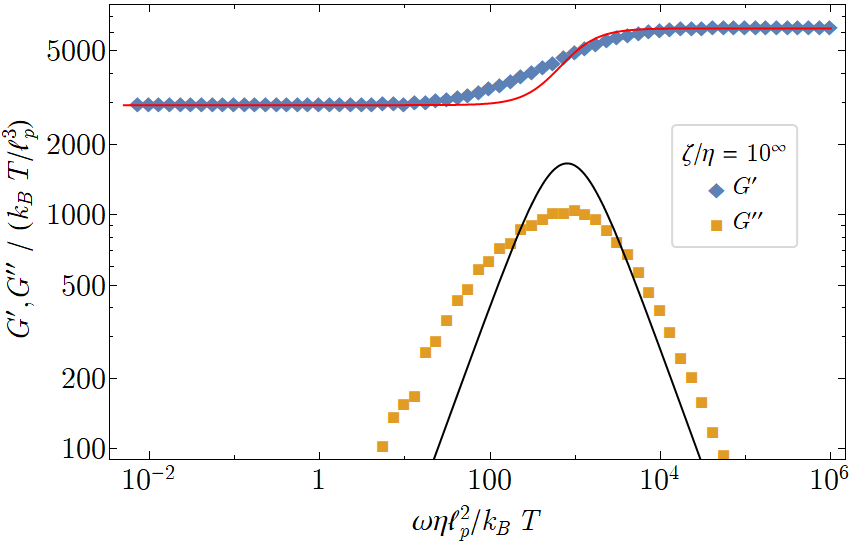}
  \caption{Storage and loss modulus of a network with permanent
    crosslinks. The solid lines represent a fit to the model of
    Eq.~(\ref{eq:model}).}
\label{fig:mod_perm}
\end{figure}

We first discuss irreversible cross-links, i.e. the limit
$\zeta/\eta\to\infty$. The measured complex shear modulus
$G(\omega)$ of the network is depicted in
Fig.~\ref{fig:mod_perm} over eight orders of magnitude in frequency.

Comparing with data from the literature one immediately recognizes the
absence of the typical high-frequency branch $G\sim
(i\omega)^{3/4}$. The reason for this is the coarse-graining procedure
intrinsic in our simulation method. The scaling with $3/4$ derives
from the competition between driving frequency and time-scale of
relaxation of bending modes with wavelength below the inter-crosslink
distance. In our simulation all these modes are assumed equilibrated,
thus no such competition exists. As explained in the introduction,
this simplification allows to increase the time-scale of the
simulation to put the emphasis on low-frequency phenomena, like
crosslink binding.

The storage modulus shows two plateaus with a transition region at
$\omega_c \approx 10^3$. Associated with this transition is a maximum in
the loss modulus $G''$. In the low-frequency plateau the loss modulus
scales as $G''\sim \omega$ and as $G''\sim \omega^{-1}$ in the
high-frequency plateau.

The crossover frequency $\omega_c$ corresponds to typical time-scales
on the length-scale of the inter-crosslink distance
$l_c\approx 0.1l_p$. For bending modes this time-scale is given by
$1/\tau_b\sim k_BTl_p/l_c^3\eta\sim 10^3$ in the units used in the
figure. The typical time-scale for stretching modes is similar,
$1/\tau_s\sim EA/l_c\eta\sim 3\cdot10^3$.

As the shear flow of the fluid mainly couples to the stretching modes,
the high-frequency response is dominated by filament stretching, when
viscous stresses force the filaments to follow the fluid flow. At
lower frequencies these stretching modes can relax and the filaments
deform mainly via bending. We have checked, by running additional
simulations with modified stretching and bending stiffness that the
modulus is indeed dominated by stretching at high frequencies and by
bending at low frequencies \cite{fischer}. \xx{In other words, the
  real part of the modulus in the high-frequency plateau is
  proportional to the stretching stiffness of the filaments, while the
  modulus in the lower plateau is proportional to the bending
  stiffness.}

A simple harmonic one-degree of freedom model can reproduce this
behavior: consider a particle coupled to two springs with spring
constants $k_1$ and $k_2$, respectively.  One of the springs is driven
by an external force that periodically changes the length $L(t)$. The
particle itself is coupled viscously to this force via $\dot L$. The
equation of motion thus reads
\begin{eqnarray}\label{eq:model}
  \eta (\dot x-\dot L) = k_2(L-x)-k_1x
\end{eqnarray}
which is easily solved for $x(t)$ assuming $L(t)=L_0e^{i\omega t}$.
Real and imaginary parts of the amplitude of oscillation are
reproduced as solid lines in Fig.~\ref{fig:mod_perm}. At high
frequencies, above $1/\tau_1=(k_1+k_2)/\eta$, the oscillation
amplitude reaches a plateau at $\Re x = L_0$ ($\Re x$ being the real
part of $x$), thus the particle follows the external driving with the
same amplitude. Both springs contribute to the response. For the
network this corresponds to the stretching dominated high-frequency
plateau. Lowering the frequency below $1/\tau_1$ the particle can
relax from the high amplitudes and $\Re x$ is reduced. Finally, at
small frequencies below $1/\tau_2=\sqrt{k_1k_2}/\eta$, the plateau
reaches the lower asymptotic value $\Re x = L_0k_2/(k_1+k_2)$. In this
limit the particle relaxes such that the load on the spring $k_1$ is
reduced. In the network this corresponds to the relaxation of the
stretching deformations such that only bending deformations remain. It
is clear from the figure that the peak width from the network
simulations is broader than the single Maxwellian peak from the simple
toy model. This is to be expected given a broad spectrum of relaxation
times in the network, as compared with the two times-scales $\tau_1$
and $\tau_2$ in the model.

We conclude that the linear response of the network to an imposed
frequency dependent strain can be divided into 3 regimes. For the
smallest frequencies all excitations are allowed to relax, giving rise
to a constant storage modulus whose value is dominated by bending
modes.  For an intermediate range of frequencies, corresponding to
typical frequencies of the spectrum of the network, the loss modulus
displays a peak, while the storage modulus increases. For the highest
frequencies beyond the spectrum of the network, the loss modulus goes
to zero, while the storage modulus is constant. Its plateau value
scales like the stretching stiffness. \xx{This interplay between
  bending and stretching, non-affine and affine response has received
  a lot of attention recently (for a review see
  \cite{broedersz2014}). Here we are mainly concerned with the effect
  of crosslink binding, which is what we turn to in the next section.}

\section{Results: Reversible cross-links}
\begin{figure}[h!tbp]\centering
\subfigure{\includegraphics[scale=0.24]{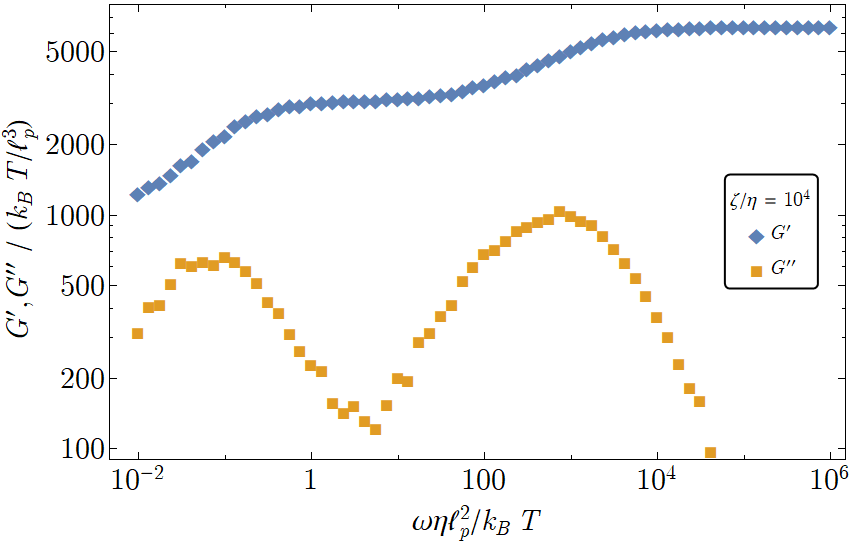}}
\caption{Storage and loss modulus as a function of frequency for a
  network with reversible crosslinks. The friction coefficient for
  crosslink sliding is taken as $\zeta/\eta=10^4$.}\label{fig:rev_mod}
\end{figure}

We now turn to the discussion of the rheology of networks with
reversible cross-links, characterized by a finite time constant
$\zeta$. As compared to the irreversible case, we expect to see a
further decrease of the storage modulus at small frequencies
corresponding to the additional relaxation mechanism of sliding
cross-links. This is indeed observed in Fig.\ref{fig:rev_mod}.

\begin{figure}[h!tbp]\centering
\subfigure{\includegraphics[scale=0.22]{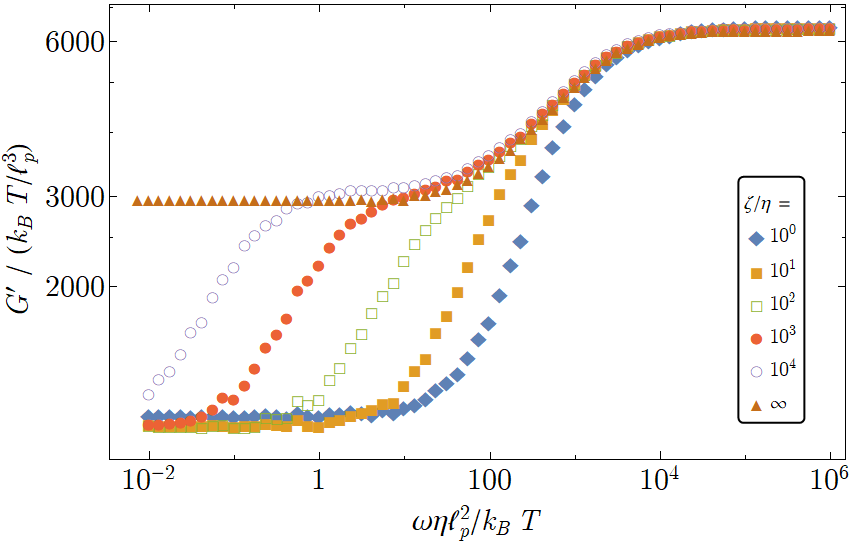}}\,\,\,
\subfigure{\includegraphics[scale=0.22]{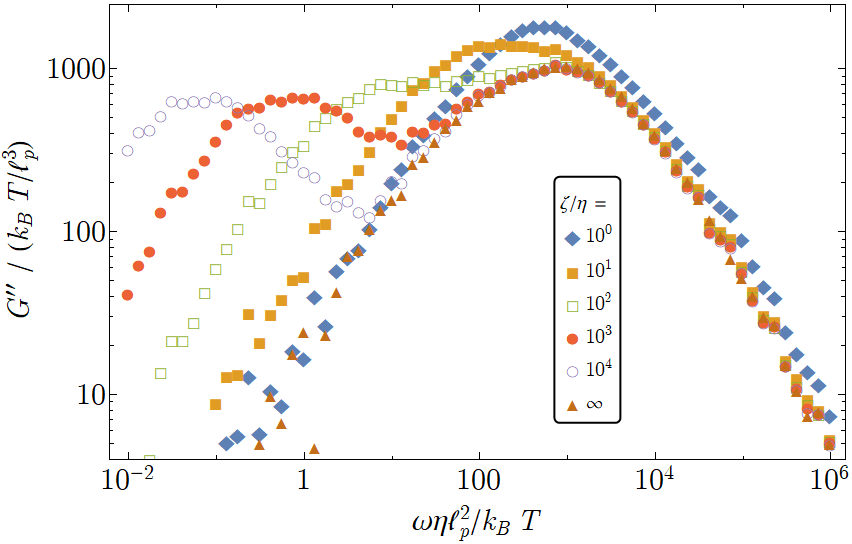}}
\caption{Storage and loss modulus as a function of frequency for
  several values of $\zeta/\eta$.}\label{fig:rev_mod_zeta}
\end{figure}

The storage modulus is seen to display three distinct plateaus: the
stretching dominated high frequency plateau, the bending dominated one
at intermediate frequencies -- both present also in the irreversible
case -- and an additional low frequency plateau (also see
Fig.~\ref{fig:rev_mod_zeta}). The latter is finite, because the
cross-links only slide along the filament and do not unbind. Also the
end-links are assumed to be non-sliding. We expect this plateau to
vanish, if the cross-links also unbind, so that a complete relaxation
of the network becomes possible.  The loss modulus displays two
distinct peaks, one corresponding to a characteristic network
frequency (as discussed above) and the other one to the inverse
relaxation time of a cross-link.

\begin{figure}[h!tbp]\centering
\subfigure{\includegraphics[scale=0.22]{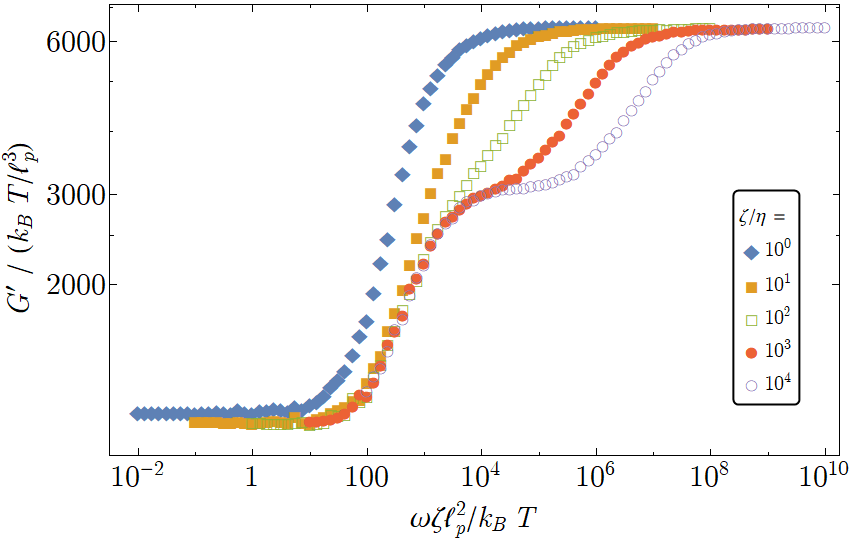}}\,\,\,
\caption{Storage modulus as a function of frequency for several values
  of $\zeta/\eta$. Frequency is scaled with $\zeta$, see text for
  details.}\label{fig:rev_mod_zeta_rescaled}
\end{figure}

The time-scale for crosslink sliding is obtained as
$1/\tau_{\rm sl}\sim k_BTl_p/\zeta l_c^3$ (wich is $\sim 0.1$ in the
figure). Thus, we expect the sliding relaxation to set in at a typical
frequency $\omega_{sl}\sim 1/\tau_{sl}\sim 1/\zeta$. In order to test
this scaling, we show results for different values of $\zeta$ in
Fig.~\ref{fig:rev_mod_zeta} . The terminal relaxation and the peak in
the loss modulus indeed shift with $\zeta$ to smaller frequencies. The
particular scaling with $1/\zeta$ is highlighted in
Fig.~\ref{fig:rev_mod_zeta_rescaled}, where the frequency axis is
rescaled by a factor $\omega_{sl}$. In particular, the low-frequency
wing is seen to scale with the crosslink time-scale $\sim\zeta$,
whenever there is a clear time-scale separation between network
relaxation processes (governed by $\eta$) and crosslink sliding
(governed by $\zeta$). In this regime the loss modulus scales as
$G''\sim \zeta\omega$ indicating viscous behavior with the viscosity
set by the crosslink sliding constant $\zeta$.

The scaling collapse at low frequencies breaks down for the case
$\zeta/\eta=1$ (blue diamonds). For this data set both viscous
processes are indistinguishable and occur on similar time-scales. This
explains the lack of scaling of these data in
Fig.~\ref{fig:rev_mod_zeta_rescaled}.

\begin{figure}[h!tbp]\centering
  \subfigure{\includegraphics[scale=0.24]{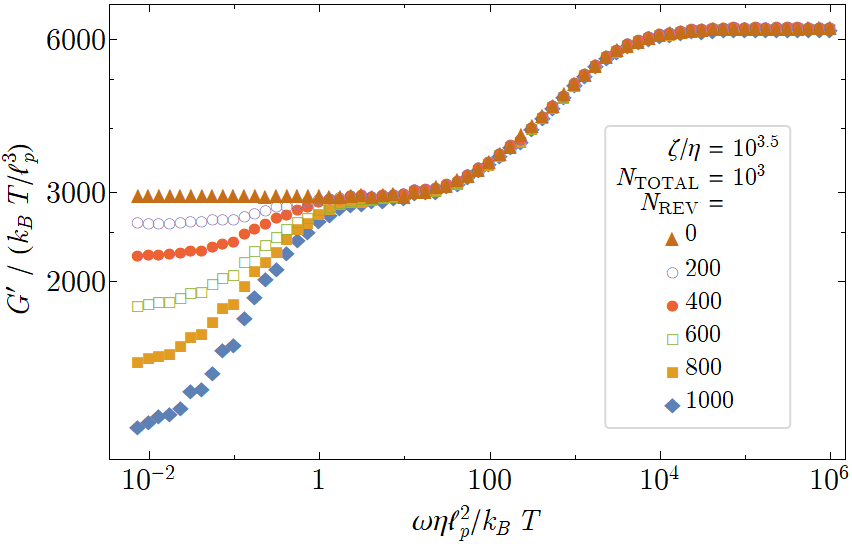}}
  \subfigure{\includegraphics[scale=0.24]{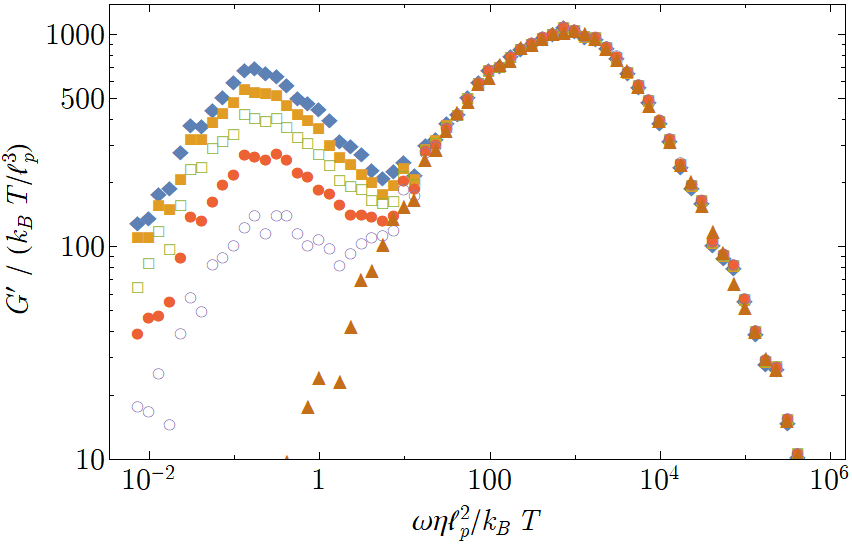}}
  \caption{Storage and loss modulus as a function of frequency for a
    network with fixed total number of crosslinks,
    $N_{\rm cl}=N_{\rm perm}+N_{\rm rev}=1000$, and changing number of
    $N_{\rm rev}$.}\label{fig:rev_perm}
\end{figure}

Further manipulation of the low-frequency plateau can be achieved by
having both, reversible and permanent, crosslinks in a
network. Fig.~\ref{fig:rev_perm} displays the modulus for networks
with a total of $N_{\rm cl}=1000$ crosslinks, a varying fraction of
which are reversible. Reducing the reversible fraction, the peak
height at low frequencies in $G''$ is reduced. We find that the peak
height scales with the number of reversible crosslinks,
$G''_{\rm peak}\sim N_{\rm rev}$. At the same time, the low-frequency
plateau in $G'$ increases. Recent calculations~\cite{Plagge2016} show
that the network should undergo a rigidity percolation transition as
the fraction of reversible crosslinks is increased.  At this point the
remaining network (made from the permanent crosslinks) becomes fluid
and is no longer able to build up forces to resist the imposed
deformation. Our data show this trend: The bending dominated plateau
of the shear modulus, which persists up to the smallest frquencies for
$N_{\rm rev}=0$, decays more and more rapidly as the fraction of
reversible crosslinks is increased.  However, the storage modulus does
not decay completely, instead a low-frequency plateau is observed in
our simulations, -- even when all crosslinks are reversible. As
discussed above, the reason for this regime is the fact that in our
simulations the crosslinks cannot fully relax and are constrained to
stay on the filaments for all times. This inhibits the full
fluidization of the sample. Other effects which can cause small, but
finite rigidity even below rigidity percolation are thermal
fluctuations~\cite{PhysRevLett.111.095503} and
prestresses~\cite{PhysRevLett.106.088301}, both of which are present
in our simulations.

In experiments on reconstituted actin networks
(e.g.~\cite{lieleg2008}) an important control parameter is the degree
of (reversible) crosslinking, measured by $R=c_x/c_a$, the ratio of
crosslink to actin monomer density.  We implement this parameter by
changing the number of connections per filament. In the above
discussed networks the average connectivity (crosslinks $N_c=1000$ per
filament $N_f=300$) was $n=2N_c/N_f=6.67$. We now reduce this number
down to $N_c=400$, or $n=2.67$ and take all crosslinks to be
reversible.

\begin{figure}[h!tbp]\centering
\subfigure{\includegraphics[scale=0.24]{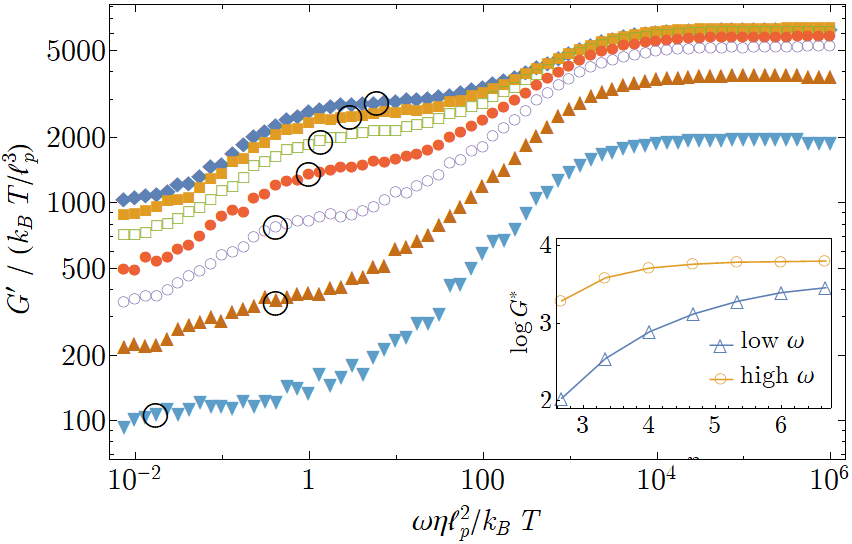}}
\subfigure{\includegraphics[scale=0.24]{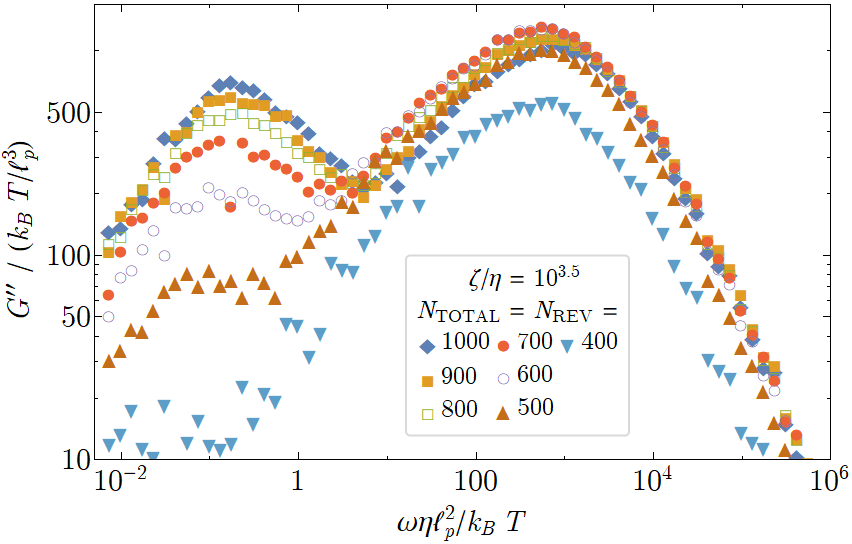}}
\caption{Storage and loss modulus as a function of frequency for
  several values of $N_{\rm rev}= 400, \ldots,1000$. Inset: plateau
  modulus as a function of crosslinks per filament $n$, extracted from
  the high-frequency (stretching-dominated) and the
  intermediate-frequency (bending-dominated) plateau of the storage
  modulus. The latter value is illustrated by the open black
  circles.}\label{fig:fewer_xlinks}
\end{figure}

The resulting storage and loss moduli are shown in
Fig.~\ref{fig:fewer_xlinks}. One observes that here the entire
frequency domain is affected by the change of $N_c$ in contrast to the
previous scenario (Fig.~\ref{fig:rev_perm}). At high frequencies, the
reversible crosslinks behave just like permanent links and the network
feels the reduced connectivity of the filaments. The reason for the
reduction of the modulus is that with fewer crosslinks, filament
segments become longer and therefore softer.

The low-frequency bending-dominated plateau is stronger affected than
the high-frequency stretching-dominated plateau, because bending
stiffness is more sensitive to segment length $l_c$,
$k_b\sim 1/l_c^3$, as compared to the stretching stiffness
$k_s\sim 1/l_c$. This difference is emphasized in the inset of
Fig.~\ref{fig:fewer_xlinks}a, where we plot the bending-dominated
plateau, $G^*$, together with the stretching dominated plateau as a
function of connectivity.  Similar results have been obtained by
Huisman {\it et al.}\cite{PhysRevLett.106.088301} in the context of an
athermal model. Parameters are slightly different, however. In our
simulations we go closer towards the percolation threshold, which also
makes the system more likely to experience finite-size effects.

The low-frequency peak in the loss modulus is strongly degraded when
the connectivity is lowered, whereas the high-frequency peak remains
nearly unchanged for most of the parameter range investigated.  For
the lowest connectivity per filament, $n=2.67$, hardly any relaxation
due to sliding crosslinks occurs. This is not surprising: Two
crosslinks at the two ends of a filament are always present and are
not allowed to slide, so that effectively less than one crosslink per
filament contributes to relaxation.  The low frequency peak in the
loss modulus is no longer detectable and the bending plateau in the
storage modulus is replaced by a finite, but small
slope~\cite{lieleg2007b,PhysRevLett.99.158105}.

. A similar
slope is observed in many epxeriments. It hints at a broadening of the
relaxation spectrum when approaching the rigidity threshold.

\begin{figure}[h!tbp]\centering
\subfigure{\includegraphics[scale=0.25]{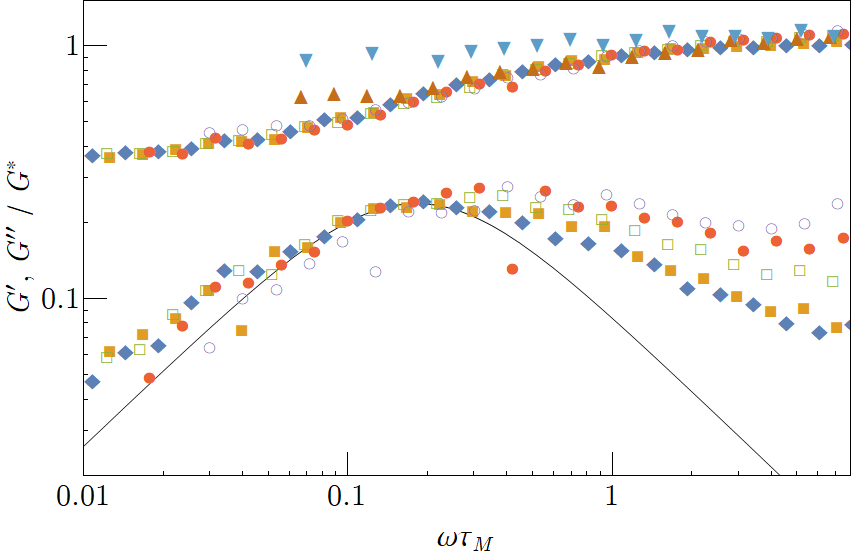}}
\caption{Storage and loss modulus (data taken from
  Fig.\ref{fig:fewer_xlinks}) as a function of frequency and rescaled
  by the plateau modulus $G^*$ (see inset of
  Fig.\ref{fig:fewer_xlinks}a) and the associated Maxwell time-scale
  $\tau_M = \zeta/G^*l_p$. Colorcode as in
  Fig.\ref{fig:fewer_xlinks}. The solid line compares with the full
  functional form of a Maxwell model
  $\sim \omega\tau/((\omega\tau)^2+1)$.}\label{fig:rescaling}
\end{figure}

The low frequency behaviour of the loss and storage moduli can be
described approximately by a Maxwell model. We use the plateau
modulus, $G^*$, depicted in the inset of Fig.~\ref{fig:fewer_xlinks}a,
as energy scale and the associated ``Maxwell'' time-scale
$\tau_M=\zeta/G^*l_p$, to rescale both, loss and storage moduli. This
rescaling, shown in Fig.~\ref{fig:rescaling}, works quite well if the
network is not too close to the percolation threshold. It is apparent
from the scaling plot for the storage modulus (upper part of
Fig.~\ref{fig:rescaling}), that the two networks closest to
percolation ($N_{\rm rev}=400,500$) hardly show any terminal
relaxation at frequencies $\omega\tau_M < 1$. In these networks there
are only few reversible crosslinks per filament and effective filament
length is short. Stress relaxation is therefore governed by filament
ends, where we have implemented permanent crosslinks that (as
explained above) are not allowed to unbind. For clarity, we have
therefore removed these two networks, when rescaling the loss modulus
(lower part of Fig.~\ref{fig:rescaling}). In the remaining networks
the loss modulus shows nice data collapse in the left wing of the
peak.  This region is dominated by crosslink (un-)binding. The absence
of scaling in the right wing is due to the transition into the second
peak of $G''$. It indicates the gradual disappearence of the
low-frequency peak within the wing of the high-frequency peak.

As compared with the functional form of a Maxwell model (solid line)
the actual peak is broader, quite similar to what is obtained for the
high-frequency peak in Fig.~\ref{fig:mod_perm}.  Theoretical
calculations~\cite{Plagge2016} show that this may be due to network
randomness, e.g. binding angles or local mesh-sizes.

\section{Conclusion and Outlook}

We have shown that the frequency-dependent elasticity of cross-linked
biopolymer networks depends strongly on the dynamics of the
crosslinks.  In our model, we consider thermal as well as forced
unbinding of the crosslinks in the periodic potential of a filament.
In response to an applied strain, the crosslinks diffuse along the
filaments, thereby partially relaxing stress. If the frequency scale
of crosslink motion is sufficiently small as compared to the
characteristic energies of the network, a distinct peak appears in the
loss modulus at about the sliding frequency. The storage modulus is
reduced correspondingly and displays an additional relaxation from the
bending dominated plateau, which for permanent crosslinks extends down
to zero frequency. The additional relaxation at the smallest
frequencies can be controlled by the relative weight of mobile to
permanent crosslinks. The observed softening of the network with
increasing fraction of reversible cross-links, indicates the loss of
shear rigidity which, however, is not complete, since the cross-links
cannot completly detach from the filaments.  At high frequencies
reversible and permanent crosslinks are indistinguishable, only the
overall connectivity determines the moduli.

Several extensions of our model are possible. Finite rates for the
crosslinks to detach and re-attach should be included in a more
realistic model of reversible cross-linking. These processes would
introduce another timescale and presumably give rise to
complete stress realaxation at the lowest frequencies.

So far we have only considered mobile passive crosslinks and focused
on their effects on stress relaxation close to quilibrium.  A
straightforward extension of our work are motors, modeled similar to
reversible crosslinks but equipped with an active velocity.  Motor
activity is known to drive the system away from thermal equilibrium
and a simple extension of our model would allow to study stress
relaxation in an active network, which is highly relevant for
biological networks as well as of fundamental interest as a model
system for nonequilibrium dynamics.

\begin{acknowledgments}
  We acknowledge financial support by the German Science Foundation
  via the Emmy Noether program (He 6322/1-1) as well as the SFB 937
  (projects A1, A16).
\end{acknowledgments}
 
\bibliography{ref,reversible}

\end{document}